\documentclass[]{aa}
\usepackage[dvips]{graphicx}
\usepackage[comma,authoryear]{natbib}
\usepackage[dvips]{graphicx}
\usepackage[english]{babel}
\usepackage[latin1]{inputenc}
\usepackage{latexsym}
\usepackage{amssymb}
\bibpunct{(}{)}{,}{a}{}{,}

\newcommand{\eg}{\textit{e.g.}}
\newcommand{\ie}{\textit{i.e.}}
\renewcommand{\l}{\ell}

\begin{document}
   \title{Mode identification in rapidly rotating stars}

   \author{D. R. Reese\inst{1} \and M. J. Thompson\inst{1} \and
           K. B. MacGregor\inst{2} \and S. Jackson\inst{2} \and
           A. Skumanich\inst{2} \and T. S. Metcalfe\inst{2}}

   \institute{
     Department of Applied Mathematics, University of Sheffield, Hicks
     Building, Hounsfield Road, Sheffield S3 7RH, UK
     \email{D.Reese@sheffield.ac.uk}
   \and
     High Altitude Observatory, National Center for Atmospheric Research,
     Boulder, CO 80307, USA
   }

   \date{Received 20 February 2009 / Accepted 24 Mai 2009}

  \abstract
   {Recent calculations of pulsation modes in rapidly rotating polytropic models
    and models based on the Self-Consistent Field method (MacGregor et al. 2007)
    have shown that the frequency spectrum of low degree pulsation modes can be
    described by an empirical formula similar to Tassoul's asymptotic formula
    (Tassoul 1980), provided that the underlying rotation profile is not too
    differential (Lignières \& Georgeot 2008, Reese et al. 2009).}
   {Given the simplicity of this asymptotic formula, we investigate whether it
    can provide a means by which to identify pulsation modes in rapidly rotating
    stars.}
   {We develop a new mode identification scheme which consists in scanning a 
    multidimensional parameter space for the formula coefficients which yield
    the best-fitting asymptotic spectra.  This mode identification scheme is
    then tested on artificial spectra based on the asymptotic formula, on random
    frequencies and on spectra based on full numerical eigenmode calculations
    for which the mode identification is known beforehand.  We also investigate
    the effects of adding random frequencies to mimic the effects of chaotic
    modes which are also expected to show up in such stars (Lignières \&
    Georgeot 2008).}
   {In the absence of chaotic modes, it is possible to accurately find a correct
    mode identification for most of the observed frequencies provided these
    frequencies are sufficiently close to their asymptotic values.  The addition
    of random frequencies can very quickly become problematic and hinder correct
    mode identification.  Modifying the mode identification scheme to reject the
    worst fitting modes can bring some improvement but the results still
    remain poorer than in the case without chaotic modes.}
   {}


   \maketitle
%

\section{Introduction}

Many stars with intermediate or high masses are rapid rotators \citep[\eg][and
references therein]{Reese2008a}.  Rapid rotation causes a number of additional
physical phenomena which make it much more difficult to model the structure and
evolution of these stars.  These include centrifugal deformation, gravity
darkening, baroclinic flows, various forms of turbulence and transport phenomena
\citep[\eg][]{Rieutord2006b}.  Much theoretical work has gone into modelling
these stars \citep[\eg][]{Meynet1997, Roxburgh2004, Roxburgh2006, Jackson2005,
MacGregor2007, Rieutord2006, Espinosa2007}. Naturally, such models are subject
to uncertainties and therefore require observational constraints. 
Asteroseismology, the study of stellar pulsations, is currently the best way to
probe the internal structure of stars and therefore to constrain such models.  A
number of recent works have therefore focused on the effects of rapid rotation,
and in particular stellar deformation, on stellar pulsations.  For acoustic
modes, these include studies based on full eigenmode calculations
\citep{Espinosa2004, Lovekin2008, Lovekin2009, Lignieres2006, Reese2006,
Reese2009} and studies based on ray dynamics \citep{Lignieres2008,
Lignieres2009}.  There are also a number of works on other types of pulsation
modes.  On the observational side, the CoRoT mission is providing stellar
pulsation data with unprecedented accuracy.  However, in order to exploit such
data, it is necessary to correctly match theoretically calculated pulsation
modes with observed ones.  This process is known as mode identification.

Until now, it has been very difficult to identify pulsation modes in rapidly
rotating stars \citep[\eg][]{Goupil2005}.  This is because mode identification
requires a proper understanding of the frequency spectrum of these stars.  Such
an understanding has only been reached recently for acoustic modes. Using ray
dynamics, \citet{Lignieres2008} and \citet{Lignieres2009} recently showed that
the acoustic spectrum of a rapidly rotating star is a superposition of spectra
from different classes of modes, each with their own frequency organisation. The
main classes are island, chaotic, whispering gallery modes and modes
corresponding to a periodic orbit of period 6.  Of particular interest are the
island modes.  These are the most visible of the regular modes since they are
the rotating counterparts to modes with low $\l-|m|$ values.  Their frequency
organisation has been studied in \citet{Lignieres2006,Lignieres2008,
Lignieres2009} and \citet{Reese2008a} for polytropic models and in
\citet{Reese2009} for models based on the Self-Consistent Field method
\citep{Jackson2005, MacGregor2007}.  A new asymptotic formula, similar to
Tassoul's formula \citep{Tassoul1980}, was derived involving a new set of
quantum numbers based on the geometry of these modes.  This naturally raises the
question as to whether such a formula can be used to identify pulsation modes.

In order to address this question, we develop a new mode identification scheme
which is described in section~2.  In section 3, we run an initial series of
tests on the mode identification scheme using frequencies based on the
asymptotic formula, random frequencies and numerical frequencies based on full
2D eigenmode calculations.  This is then followed by tests using composite
spectra in which random frequencies have been added to a set of numerical
frequencies. The final section concludes by discussing the results.

\section{A new mode identification scheme}

Various methods for identifying pulsation modes or detecting underlying
regularities in frequency spectra have been invented over the past few years. 
For instance, \citet{Breger1999} and \citet{Breger2009} have worked with
histograms of frequency differences in order to interpret pulsation spectra of
$\delta$ Scuti stars.  Other similar techniques include calculating the
autocorrelation function or the Fourier transform of the power spectrum to
identify the large frequency separation in solar-like pulsators \citep[\eg][and
references therein]{Chaplin2008}. These procedures yield information on the
structure of the frequency spectrum rather than a detailed identification for
each pulsation mode.  Another type of approach has consisted in directly
comparing the set of observed pulsation modes to numerically calculated
frequency spectra from models in a large parameter space.  Given the
computational cost involved in computing each pulsation spectrum, various
methods have been created in order to search through parameter space in an
intelligent way.  For instance, \citet{Metcalfe2003} and \citet{Charpinet2005}
have used genetic algorithms to find best matching models for white dwarfs and
sdB stars, and \citet{Bazot2008} have used a Monte Carlo Markov Chain (MCMC)
approach when studying $\alpha$ Cen A.  As opposed to other methods, this type
of approach yields potential mode identifications for each frequency in the
observed spectrum.

The mode identification scheme described here combines some of the
characteristics of the previous methods.  On the one hand, it would be nice to
get more detailed information than what is available from histograms of
frequency differences or autocorrelation functions and Fourier transforms of
power spectra.  Furthermore, such methods may fail in rapidly rotating stars
because the geometric term $-m\Omega$ due to advection of modes by rotation may
give rise to frequency spacings which are comparable to the separation between
island modes with consecutive $\tilde{n}$ values (which corresponds to half the
large frequency separation).  On the other hand, it is not currently feasible to
calculate complete pulsation spectra of rapidly rotating 2D models at each point
in a large multidimensional parameter space, even with schemes such as genetic
algorithms or the MCMC approach.  A useful compromise is therefore to use an
asymptotic formula to calculate approximate frequency spectra and to explore the
parameter space based on the coefficients of this formula in search of best
fitting spectra.

In what follows, we used the following asymptotic formula based on
\citet{Reese2009} to construct approximate frequency spectra:
\begin{equation}
      \omega = \tilde{n} \Delta_{\tilde{n}}
             + \tilde{\l}\Delta_{\tilde{\l}}
             + m^2 \Delta_{\tilde{m}}
             - m \Omega
             + \tilde{\alpha},
\end{equation}
where $\Delta_{\tilde{n}}$, $\Delta_{\tilde{\l}}$, $\Delta_{\tilde{m}}$, and
$\tilde{\alpha}$ are coefficients which depend on the stellar
structure, $\Omega$ the rotation rate and $\tilde{n}$, $\tilde{\l}$ and $m$
quantum numbers.  These quantum numbers correspond to the number of nodes along
and parallel to the underlying ray paths \citep[see, for example, Fig.~3
of][]{Reese2008c} and to the usual azimuthal order, respectively.  In what
follows, we will refer to $\tilde{n}$ as a radial order although it roughly
corresponds to twice the usual (spherical) radial order.  This formula is an
approximation which is valid at low azimuthal orders of a more complete
formula.  Given the computational efficiency of calculating a single spectrum
using the asymptotic formula, a simple scan of the parameter space, in
which the coefficients and rotation rate are treated as the parameters, is
performed rather than applying a more sophisticated search algorithm.

\begin{figure}[htbp]
\centering
\includegraphics[width=\columnwidth]{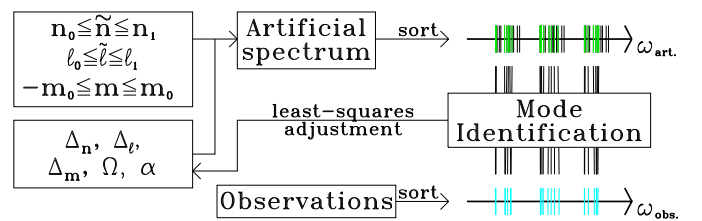}
\caption{Schematic diagram which shows how one asymptotic spectrum can be
compared with a set of observed frequencies to produce a plausible mode
identification.}
\label{fig:one_comparison}
\end{figure}

A number of preliminary choices are made before applying this method.  Suitable
ranges of $\tilde{n}$, $\tilde{\l}$ and $m$ values must be selected.  These will
determine which frequencies are calculated in the asymptotic spectra.  Also,
bounds must be set for the parameter space. Besides these choices, the set of
observed frequencies is assumed to be sorted in ascending order as this
is needed when matching these frequencies with those from the asymptotic
spectra. The following procedure is then applied to each point in the parameter
space:
\begin{enumerate}
\item An artificial spectrum based on the asymptotic formula is created
      using the values of the different parameters and the ranges chosen for
      $\tilde{n}$, $\tilde{\l}$ and $m$.
\item The artificial spectrum is sorted in ascending order using a
      heapsort method.
\item The observed frequencies are matched to neighbouring asymptotic
      frequencies using dichotomy.  When two or more observed frequencies are
      nearest to the same artificial frequency, only the first one is matched to
      the frequency, the others being matched to following frequencies.  In some
      cases this can produce sub-optimal solutions.
\item This match between the artificial and observed frequencies yields a mode
      identification.  Based on this mode identification, the formula
      coefficients are then recalculated through a least-squares minimisation of
      the standard deviation between the artificial frequencies and the observed
      ones.
\end{enumerate}
Figure~\ref{fig:one_comparison} illustrates these different steps.

The search domain is in terms of the following parameters rather than
the original coefficients from the asymptotic formula:
\begin{equation}
      \Delta_{\tilde{n}}, \quad
      \frac{\Delta_{\tilde{\l}}}{\Delta_{\tilde{n}}}, \quad
      \frac{\Delta_{\tilde{m}}}{\Delta_{\tilde{n}}}, \quad
      \frac{\Omega}{\Delta_{\tilde{n}}}, \quad
      \frac{\alpha}{\Delta_{\tilde{n}}}.
\label{eq:asymptotic}
\end{equation}
The advantage of working with these parameters is that the artificial spectrum
does not need to be sorted again when the value of $\Delta_{\tilde{n}}$ or
$\frac{\alpha}{\Delta_{\tilde{n}}}$ is modified.

It is important to note that the synthetic spectra are periodic with a period
equal to $\Delta_{\tilde{n}}$, except for the ends where frequencies will be
missing due to the cutoff in $\tilde{n}$ values.  As a result of this, it is
only necessary to test the parameter $\frac{\alpha}{\Delta_{\tilde{n}}}$ within
a unit interval.  Choosing the wrong interval will offset all of the $\tilde{n}$
values by a fixed amount.  Since this is probably one of the least robust
parameters, it is likely that the mode identification scheme will only yield
relative rather than absolute $\tilde{n}$ values for real stars.

\section{Results}

\subsection{Initial tests}

A number of artificial spectra in which the identification of the frequencies is
known beforehand were used to test the mode identification scheme.  These
spectra included artificial frequencies which follow exactly the asymptotic
formula, random frequencies, and frequencies of low and high order modes of a
$25\,\,M_\odot$ ZAMS model rotating uniformly at $60 \%$ of the
critical rotation rate \citep[for details on the eigenmode calculations,
see][]{Reese2009}.  Results from these tests are summarised in
Table~\ref{tab:first_tests}.  In cases 1, 2, 4 and 5, $50$ frequencies
were randomly selected with $\tilde{\l}$ values equal to $0$ or $1$ and $m$
values between $-3$ and $3$.  The $\tilde{n}$ values are given in the fourth
column of Table~\ref{tab:first_tests}.  For cases 1, 2 and 4, this corresponds
to 50 out of 84 possible modes, whereas there are 224 possible modes for the
last case.  Case 3 corresponds to a set of $50$ frequencies with random
values in the same frequency range as case 2.  As such, these frequencies have
no corresponding identifications.  The third column gives the standard
deviation $\delta \omega$ between the frequencies and their asymptotic
approximations, normalised by $\Delta_{\tilde{n}}$.  This deviation is defined
as follows:
\begin{equation}
\delta \omega = \sqrt{\frac{1}{N} \sum_{i=1}^{N} \left(\omega_i -
                \omega_i^{\mathrm{asymp.}} \right)^2},
\end{equation}
where $N$ is the number of observed modes, $\omega_i$ the ``observed''
frequencies and $\omega_i^{\mathrm{asymp.}}$ their asymptotic approximations,
as based on Eq.~(\ref{eq:asymptotic}).  The coefficients in
Eq.~(\ref{eq:asymptotic}) can be calculated in several ways. 
\citet{Lignieres2008} give theoretical formulas for $\Delta_{\tilde{n}}$ and
$\Delta_{\tilde{\l}}$ based on travel-time integrals of underlying ray paths. 
However, similar formulas for the remaining coefficients are not currently
available.  A more heuristic approach is to calculate a set of numerical
frequencies and find the corresponding coefficients using a least-squares fit. 
This then raises the issue as to which frequencies are to be included in the
set.  In the current context, the most logical choice is the set of ``observed''
frequencies specific to each of the cases.  Indeed, the mode identification
scheme can only find the asymptotic coefficients based on the frequencies which
are available.  Furthermore, choosing these frequencies yields the lowest value
for $\delta \omega$.  This implies, however, that the asymptotic coefficients
will be different for cases 2, 4 and 5 (as can be seen in
Table~\ref{tab:bounds}) in spite of the fact that these correspond to the same
model.

\begin{table*}[htbp]
\begin{center}
\caption{Description of the ``observed'' frequencies and parameters/results from
the mode identification scheme.}
\label{tab:first_tests}
\begin{tabular}{*{7}c}
\hline
\hline
&
\multicolumn{3}{c}{\dotfill ``Observed'' frequencies \dotfill} &
\multicolumn{3}{c}{\dotfill Mode identification \dotfill} \\
Case &
Type &
$\displaystyle \frac{\delta \omega}{\Delta_{\tilde{n}}}$ &
$\tilde{n}$ &
$\tilde{n}$ &
Average success &
$\displaystyle \frac{\delta \omega}{\Delta_{\tilde{n}}}$ \\
\hline
1 & Asymptotic &                    0 & 10-25 & 10-25 & 77.4 \% & $3.7 \times 10^{-3}$ \\
2 & Numerical  & $2.4 \times 10^{-2}$ & 15-20 & 10-25 & 17.5 \% & $1.2 \times 10^{-2}$ \\
3 & Random     &         -            &   -   & 10-25 &    -    & $2.1 \times 10^{-2}$ \\
4 & Numerical  & $2.8 \times 10^{-3}$ & 45-50 & 40-55 & 65.9 \% & $5.4 \times 10^{-3}$ \\
5 & Numerical  & $7.9 \times 10^{-3}$ & 40-55 & 35-60 &  73.9 \% & $7.3 \times 10^{-3}$ \\
\hline
\end{tabular}
\end{center}
\begin{list}{}{}
\item[] Characteristics of the ``observed'' frequency spectra (columns 2-4), of
the $\tilde{n}$ values used in the mode identification scheme (column 5) and of
the results (columns 6-7).  The observed frequencies from cases 2, 4 and 5 come
from a $25\,\,M_\odot$ model, uniformly rotating at $60 \%$ of the critical
rotation rate.
\end{list}
\end{table*}

The mode identification scheme was then applied using the same $\tilde{\l}$ and
$m$ ranges.  The range on $\tilde{n}$ used by the scheme was chosen to be larger
than the range used to generate the observations (except for the first case), as
can be seen in the fifth column of Table~\ref{tab:first_tests}.  The bounds of
the parameter space are given in Table~\ref{tab:bounds} along with the values of
the parameters corresponding to the exact solution.  The parameter space was
discretised using $50$ uniformly distributed grid points in each direction, thus
yielding a total of $3.125 \times 10^8$ combinations, except for case 5
where $100$ uniformly distributed points were used in each direction.  The
average computational time was approximately 1 hour on a single PC for
cases 1-4, and around 20 hours for case~5.

\begin{table}[htbp]
\begin{center}
\caption{Bounds to parameter space and solutions.}
\label{tab:bounds}
\begin{tabular}{*{6}c}
\hline
\hline
Case(s) &
$\begin{array}{c} \Delta_{\tilde{n}} \\ \mbox{(in $\mu$Hz)} \end{array}$ &
$\displaystyle \frac{\Delta_{\tilde{\l}}}{\Delta_{\tilde{n}}}$ &
$\displaystyle \frac{\Delta_{\tilde{m}}}{\Delta_{\tilde{n}}}$ &
$\displaystyle \frac{\Omega}{\Delta_{\tilde{n}}}$ &
$\displaystyle \frac{\alpha}{\Delta_{\tilde{n}}}$ \\
\hline
\multicolumn{6}{c}{Bounds to parameter space} \\
\hline
1   & 11.5-19.2 & 0.5-0.9 & 0.01-0.05 & 0.6-1.0 & 2.5-3.5 \\
2,3 & 9.6-19.2  & 0.5-0.9 & 0.00-0.05 & 0.8-1.2 & 2.8-3.8 \\
4,5 & 9.6-19.2  & 0.5-0.9 & 0.00-0.05 & 0.8-1.2 & 1.0-2.0 \\
\hline
\multicolumn{6}{c}{Solutions} \\
\hline
1$^\star$ & 17.26 & 0.660 & 0.0288 & 0.827 & 2.92 \\
2 & 15.25 & 0.767 & 0.0205 & 0.955 & 3.35 \\
4 & 16.01 & 0.755 & 0.0076 & 0.919 & 1.64 \\
5 & 16.01 & 0.758 & 0.0074 & 0.921 & 1.65 \\
\hline
\end{tabular}
\end{center}
\begin{list}{}{}
\item[] Bounds of the parameter space used in the mode identification scheme
and corresponding exact solutions.  These bounds were chosen so as to include
the solutions. 
\item[] $^\star$These parameters correspond to a polytropic model with a
polytropic index of $3$ and the same mass and equatorial radius as the model
used in cases 2, 4 and 5 (\ie\ $M = 25 M_{\odot}$ and $R_{\mathrm{eq}} = 7.46
R_{\odot}$).
\end{list}
\end{table}

Column 6 of Table~\ref{tab:first_tests} gives the success rate in correctly
identifying modes, averaged over the 100 best solutions given by the
identification scheme.  The last column gives the lowest standard deviation
between the fitted frequencies and observations, normalised by the
$\Delta_{\tilde{n}}$ of the output solutions.

As could be expected, using frequencies which follow exactly the asymptotic
formula yields the best results.  Although the exact solution is within the
bounds of the parameter space tested, it was not actually found -- the nearest
grid points yielded mode identifications which were slightly different than the
original.

Using low order numerical frequencies does not yield good results, as can be
seen from case 2.  The reason for this failure seems straightforward. At low
radial orders, the deviations caused by avoided crossings between the
frequencies and their asymptotic values can be substantial.  As a result, the
mode identification scheme found erroneous identifications which actually led to
closer fits to the numerical frequencies than the original identifications. This
point is further confirmed by case 3, in which the mode identification
scheme is able to reproduce a set of random frequencies with no underlying
regularities with a standard deviation equal to $0.02 \Delta_{\tilde{n}}$.  This
is marginally worse than the standard deviation obtained in case 2.

\begin{figure*}[htbp]
\begin{minipage}{0.65\textwidth}
\includegraphics[width=\textwidth]{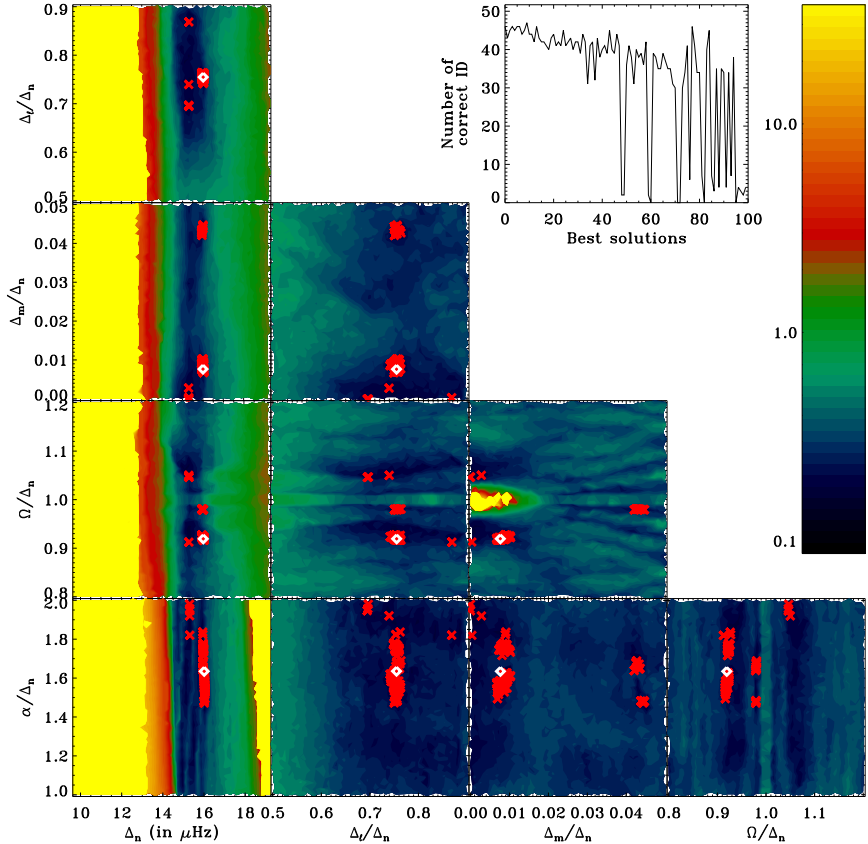}
\end{minipage} \hfill
\begin{minipage}{0.34\textwidth}
\caption{Plots which show the standard deviation, $\delta \omega$, in
parameter space for case~4 of Table~\ref{tab:first_tests}. The ten plots
arranged in triangular form correspond to different pairs of parameters.  For
each pair, the plot shows $\delta \omega$ as a function of the two parameters,
the remaining parameters having been optimised over the search domain. 
Superimposed on these plots are the 100 best solutions (as shown by the red
crosses) and the exact solution (indicated by the white diamond).  The plot in
the upper right corner shows the number of correctly identified pulsation modes
for the 100 best solutions.}
\label{fig:opt_nobs50_high_n}
\end{minipage}
\end{figure*}

Going to higher radial orders substantially reduces the deviations between the
numerical frequencies and their asymptotic values.  This can be seen by
comparing the standard deviations of cases 2 and 4 (see 3$^{\mathrm{rd}}$ column
of Table~\ref{tab:first_tests}) which shows a tenfold decrease at higher radial
orders.  This leads to good results when applying the mode identification
scheme.  Figure~\ref{fig:opt_nobs50_high_n} shows a series of ten plots arranged
in triangular form which give an idea of the accuracy of the fits in parameter
space. Each plot corresponds to different pairs of parameters, which we shall
generically denote as $p_1$ and $p_2$.  The plots show the standard deviation,
$\delta \omega$, as a function of $p_1$ and $p_2$, the remaining variables being
optimised over the parameter space. The colour bar on the righthand side
indicates the meaning of the different colour levels.  Since these plots are
based on the parameters after the least-squares adjustment, the different
positions in each plots are actually bins in which only the best solution is
retained.  Using the adjusted parameters also means that some regions will be
avoided, and these are indicated in yellow.

Superimposed on these plots are the positions of the 100 best solutions, as
represented by the red crosses.  As can be expected, these crosses are
concentrated in the dark regions which correspond to the best fits.   The white
diamonds show the exact solution.  The plot in the upper right corner, besides
the colour bar, indicates the number of modes correctly identified.  As can be
seen from these plots, the correct solution has a basin around it which
attracts most of the 100 best-fitting solutions.  Some other secondary basins
around other solutions also appear and attract a few of the best-fitting
solutions.

The last case in Table~\ref{tab:first_tests} tests the effects of using a larger
range of $\tilde{n}$ values and therefore a sparser set of observed
frequencies.  In order to obtain good results, it was necessary to use a
finer grid, hence the reason why $100$ rather than $50$ points were used in each
direction.  This need for a higher resolution indicates that the basin around
the exact solution is smaller, probably as a result of the larger range of $n$
values.  Once a sufficient grid resolution is chosen, then the mode
identification scheme can yield very good results in spite of the sparseness of
the set.

\begin{figure}[htbp]
\includegraphics[width=\columnwidth]{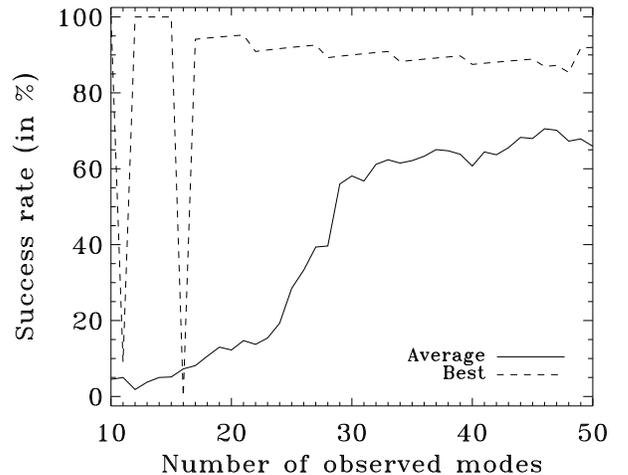}
\caption{The average success rate of the 100 best-fitting solutions (solid line)
and the success rate of the best solution (dotted line).}
\label{fig:nobs}
\end{figure}

Another interesting test consists in changing the number of observed modes. 
This has been done for case 4 of Table~\ref{tab:first_tests}.  Two different
measurements of the success rate are shown in Fig.~\ref{fig:nobs}.  The solid
line shows the average success rate of the 100 best-fitting solutions whereas
the dotted line shows that of the best solution.  As the number of
observed frequencies decreases, many additional local minima appear in parameter
space due to the lower number of constraints.  These cause the average success
rate to decrease, especially when there are fewer than $30$ modes, by attracting
an increasing number of the 100 best-fitting solutions away from the true
solution.  The dotted line, on the other hand, shows that in most cases, the
best-fitting solution still remains the true solution.  However, in some cases,
one or several other local minima with completely different mode identifications
produce solutions which fit the observations even better than the true solution,
thereby causing a dramatic drop in the accuracy of the best solution. This is
illustrated for $11$ and $16$ observed modes in Fig.~\ref{fig:nobs}.  As a
result, the best-fitting solution must be used with caution.

\subsection{Tests with additional random frequencies}

As was shown in \citet{Lignieres2009} based on geometric visibility
calculations, chaotic modes are likely to be visible in the pulsation spectrum
of rapidly rotating stars.  The frequencies of these modes do not follow an
asymptotic formula but rather a statistical distribution.  These will naturally
make mode identification more difficult as it is not possible to know  \textit{a
priori} which modes are regular and which ones are chaotic in an observed
frequency spectrum.  In order to mimic the presence of chaotic modes, we have done
a number of tests in which random frequencies were added to the
frequencies from cases 2 and 4 of Table~\ref{tab:first_tests}.

\begin{figure*}[htbp]
\includegraphics[width=\columnwidth]{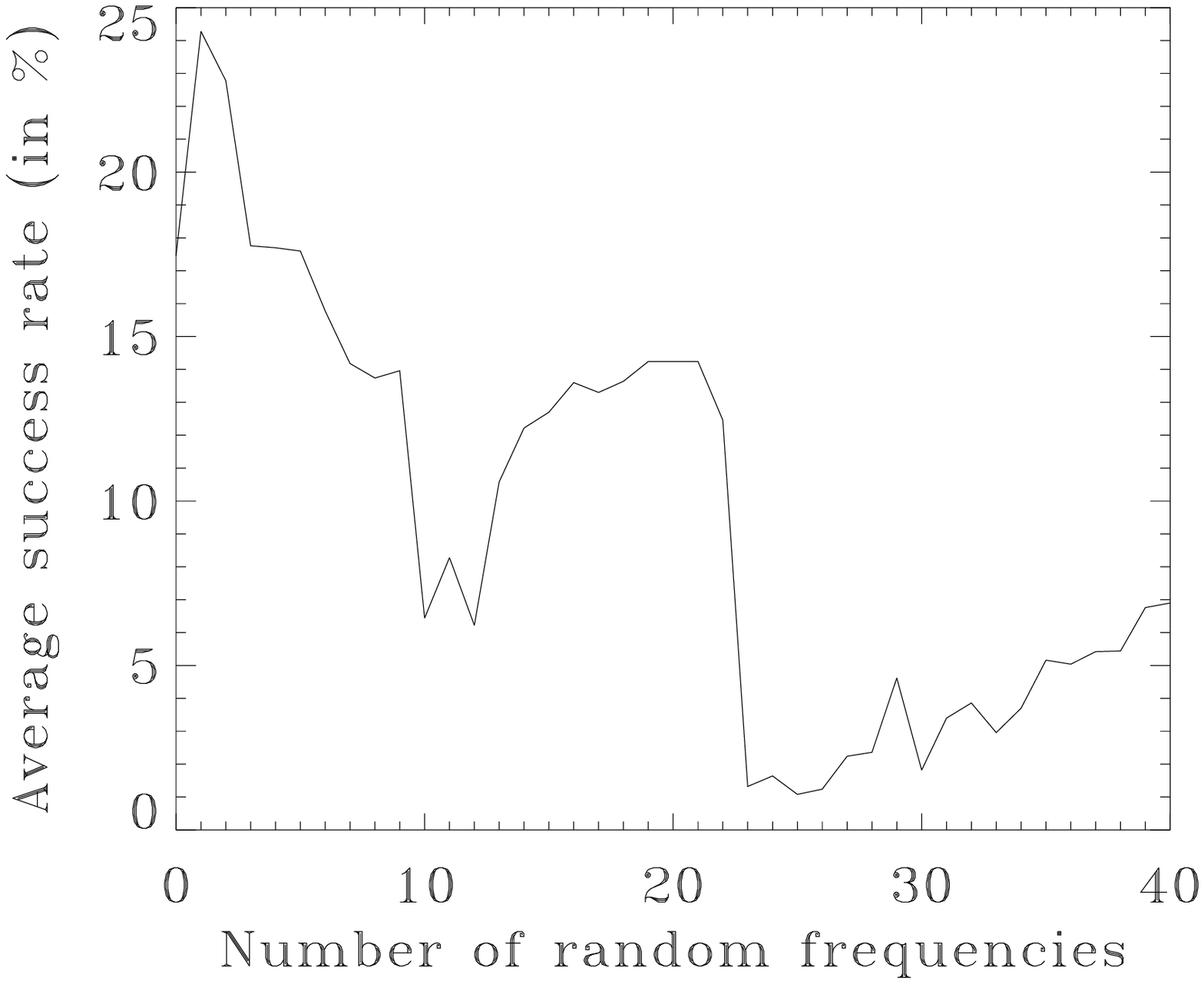} \hfill
\includegraphics[width=\columnwidth]{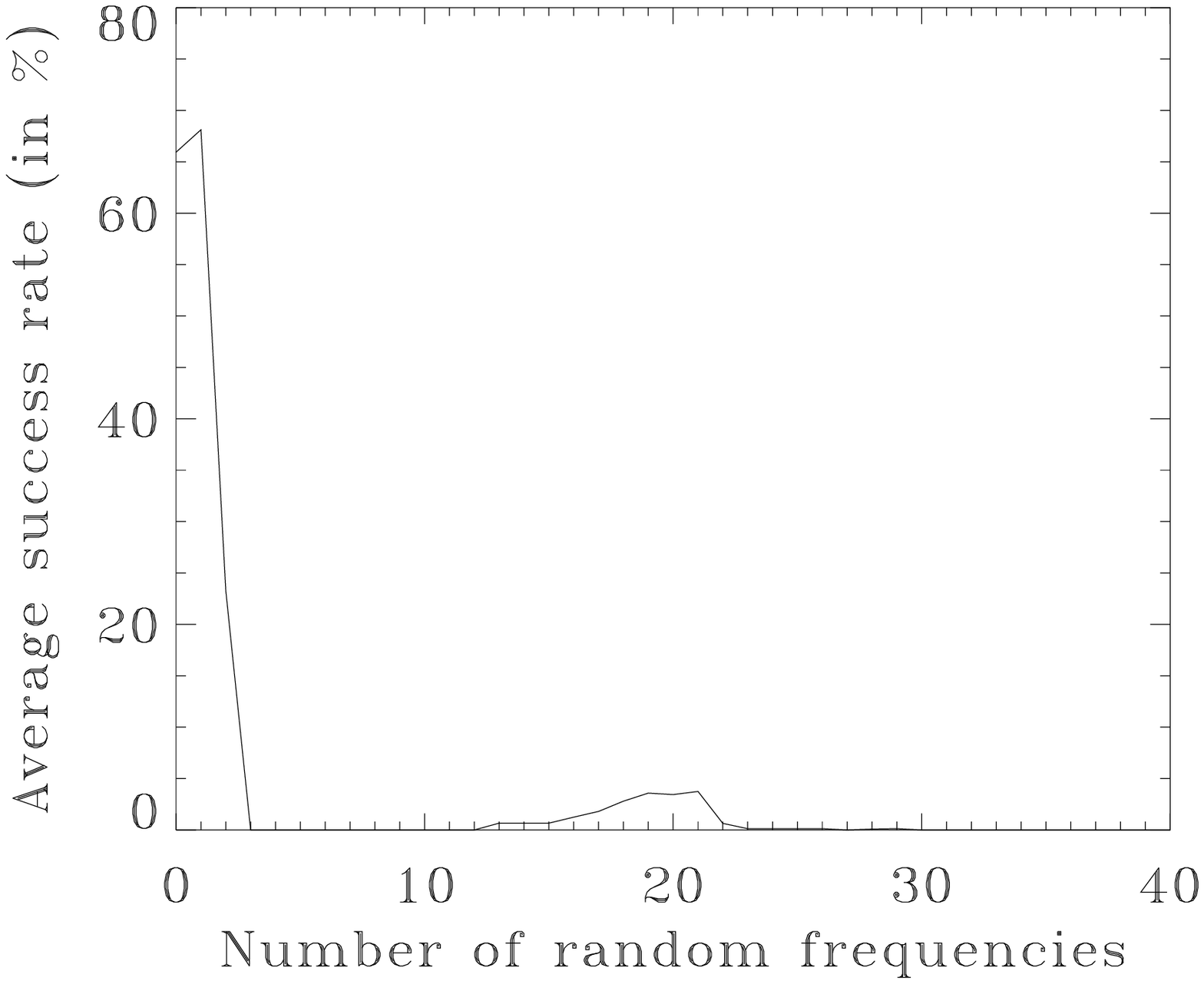}
\caption{Average success rate as a function of the number of additional random
frequencies.  The left plot corresponds to low order modes (case 2 of
Table~\ref{tab:first_tests}) and the right plot to high orders (case
4).}
\label{fig:nrand}
\end{figure*}

Figure~\ref{fig:nrand} shows two plots with the average success rate as a
function of the number of additional random frequencies, one for case 2
(left panel) and the other for case 4 (right panel).  Although low
order modes yielded worse results without random frequencies, they seem to be
less affected by the presence of random frequencies than high order modes.  The
reason why the results for high order modes are so poor seems to be that random
frequencies shift the basin of best-fitting solutions away from the true
solution.  Even a small deviation between the two can be sufficient to throw off
the identification of modes.

\begin{figure}[htbp]
\centering
\includegraphics[width=\columnwidth]{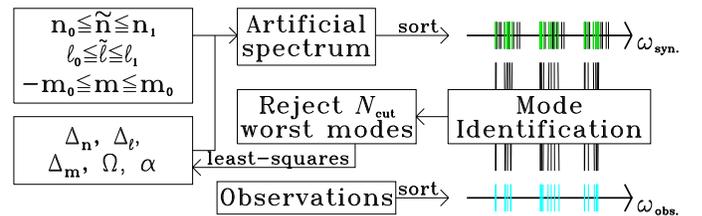}
\caption{Same diagram as Fig.~\ref{fig:one_comparison} except for an
additional step in which the $N_{\mathrm{cut}}$ worse frequencies are removed.}
\label{fig:one_comparison_ncut}
\end{figure}

One way of trying to deal with random frequencies is to add another
step in the mode identification procedure.  At each point in parameter space,
after matching the artificial frequency spectrum with the observed frequencies
(step 3 of the procedure described in Section 2), one can remove the
$N_\mathrm{cut}$ worst modes, where $N_\mathrm{cut}$ is a fixed number. 
Figure~\ref{fig:one_comparison_ncut} illustrates the mode identification
procedure with this extra step.

\begin{figure}[htbp]
\centering
\includegraphics[width=\columnwidth]{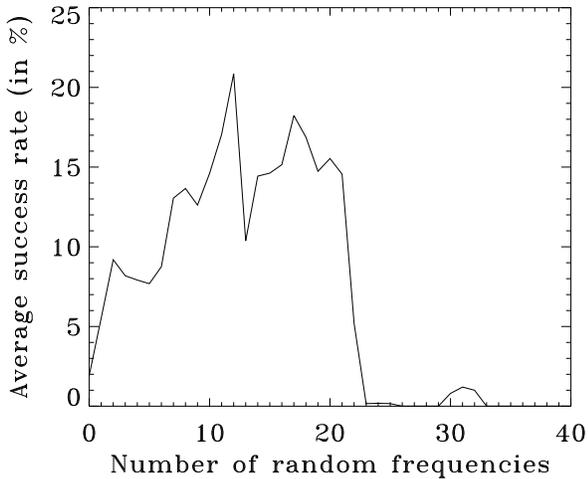}
\caption{Average success rate as a function of the number of additional random
frequencies.  The frequency sets are the same as those used in the right panel
of Fig.~\ref{fig:nrand}.  This time, the $N_{\mathrm{cut}}=20$ worst frequencies
have been rejected at each point in parameter space.  The success rate has been
calculated based on the modes which have been retained: rejecting non-random
frequencies has not been penalised.}
\label{fig:nrand_ncut}
\end{figure}

Figure~\ref{fig:nrand_ncut} shows the effects of removing $N_{\mathrm{cut}}=20$
frequencies on the average success rate as a function of the number of
additional random frequencies.  The frequency sets are the same as those used in
the right panel of Fig.~\ref{fig:nrand}.  The success rate has been calculated
only using the modes which have been retained: rejecting non-random frequencies
has not been penalised.  As can be seen, having a comparable number of random
frequencies to $N_{\mathrm{cut}}$ yields the best results.  The reason why the
results are poor when there are no random frequencies is because additional
solutions with lower standard deviations than the original solution are
appearing.  This is similar to what happened when the number of observed modes
was reduced (see Fig.~\ref{fig:nobs}), although this time the problem is worse
because the mode identification is carefully choosing the subset of frequencies
so as to reduce the standard deviation.  When random frequencies are added,
these tend to be rejected rather than the true frequencies, thereby leading to
more favourable results.  Even in the best of situations, identifying modes
remains difficult when random frequencies are present.

\section{Conclusion}

In this paper we have created and tested a new method for identifying pulsation
modes in rapidly rotating stars, based on an asymptotic formula for island
modes, the rotating counterparts to acoustic modes with low $\l-|m|$ values
\citep{Lignieres2008, Reese2009}.  This method consists in scanning a parameter
space in search of the formula coefficients which lead to the best agreement
with observations.  Results show that it is possible to correctly identify
pulsation modes provided they are sufficiently numerous and their frequencies
close enough to their asymptotic values.  Such a situation occurs for
30 or more high order modes, typically in the $\tilde{n} = 40-50$
range.  Bearing in mind that $\tilde{n}$ is roughly twice the spherical radial
order, this range is similar to that of the solar p-modes.  This method may
apply to $\delta$ Scuti stars thanks to their numerous pulsation frequencies,
but these will have to be of sufficiently high radial order.

However, when even a few random frequencies are added in order to mimic the
presence of chaotic modes, results can become quite poor.  Adding an extra step,
in which the worst modes are removed, can bring a small improvement but this
still remains insufficient to enable a reliable mode identification.  Recent
calculations by \citet{Lignieres2009} suggest that chaotic modes may be as much
as 5 times as numerous as island modes for a star at $60 \%$ of the critical
rotation rate, which is far more than what is considered here.  One way to try
to deal with this problem is to start with stars at lower rotation rates, where
a smaller fraction of the modes are chaotic.

One of the weaknesses of this method is the large number of input parameters.
These include the ranges on the different quantum numbers and the bounds on
parameter space.  In order to set these parameters correctly, prior knowledge of
the star through other observations will be needed as well as an understanding
of the relationships between the coefficients in the asymptotic formula and
fundamental stellar parameters such as the mass, age and rotation rate.  Such an
understanding can only be reached through a systematic study of the frequency
spectra of a grid of stellar models.


Potential improvements in this method are as follows.  Rather than doing a
simple scan of the parameter space, using a more sophisticated search method
like a genetic algorithm \citep[\eg][]{Metcalfe2003, Charpinet2005} or an MCMC
method  \citep[\eg][]{Bazot2008} would enable a more detailed investigation of
regions of interest while spending less time in other less important parts of the
parameter space. This would be beneficial to case 4 of
Table~\ref{tab:first_tests} as the solutions that were found were not as optimal
as the true solution, and case 5 because the basin around the true
solution is smaller.  Also, correlating mode visibilities with observed mode
amplitudes could provide supplementary constraints.  This, of course, is only
viable if a simple way to estimate mode visibilities exists.  It is then hoped
that the results from this method would provide a useful starting point for
comparing full 2D calculations to observations.

\begin{acknowledgements}
Many of the numerical calculations were carried out on Iceberg (University of
Sheffield) and on the Altix 3700 of CALMIP (``CALcul en
MIdi-Pyr{\'e}n{\'e}es''), both of which are gratefully acknowledged.  DRR
gratefully acknowledges support from the UK Science and Technology Facilities
Council through grant ST/F501796/1, and from the European Helio- and
Asteroseismology Network (HELAS), a major international collaboration funded by
the European Commission's Sixth Framework Programme.  The National Center for
Atmospheric Research is sponsored by the National Science Foundation.
\end{acknowledgements}

\bibliographystyle{aa}
\bibliography{biblio}


\end{document}